\documentclass[
    reprint,
    superscriptaddress,
    amsmath,amssymb,
    aps,
    prx,
    floatfix,
]{revtex4-2}

\usepackage{graphicx} 
\usepackage{xcolor}
\usepackage{siunitx}
\usepackage{braket}
\usepackage{bm}
\usepackage[colorlinks=true,linkcolor=blue,citecolor=blue]{hyperref}
\usepackage[capitalise]{cleveref}

\newcommand{\conc}{\mathcal{C}}

\makeatletter
\renewcommand{\fnum@figure}{\textbf{Fig.~\thefigure}}
\makeatother

\begin{document}

\title{Autonomous stabilization of remote entanglement in a cascaded quantum network}

\author{Abdullah Irfan}
\thanks{These authors contributed equally.}
\author{Kaushik Singirikonda}
\thanks{These authors contributed equally.}
\affiliation{Department of Physics, The Grainger College of Engineering, University of Illinois at Urbana-Champaign, Urbana, IL 61801, USA}
\author{Mingxing Yao}
\affiliation{Pritzker School of Molecular Engineering, University of Chicago, Chicago, IL 60637, USA}
\author{Andrew Lingenfelter}
\affiliation{Pritzker School of Molecular Engineering, University of Chicago, Chicago, IL 60637, USA}
\affiliation{Department of Physics, University of Chicago, Chicago, IL 60637, USA}
\author{Michael Mollenhauer}
\author{Xi Cao}
\affiliation{Department of Physics, The Grainger College of Engineering, University of Illinois at Urbana-Champaign, Urbana, IL 61801, USA}
\author{Aashish A. Clerk}
\affiliation{Pritzker School of Molecular Engineering, University of Chicago, Chicago, IL 60637, USA}
\author{Wolfgang Pfaff}
\email{wpfaff@illinois.edu}
\affiliation{Department of Physics, The Grainger College of Engineering, University of Illinois at Urbana-Champaign, Urbana, IL 61801, USA}
\affiliation{Materials Research Laboratory, The Grainger College of Engineering, University of Illinois at Urbana-Champaign, Urbana, IL 61801, USA}
\affiliation{Holonyak Micro and Nanotechnology Lab, The Grainger College of Engineering, University of Illinois at Urbana-Champaign, Urbana, IL 61801, USA}
\affiliation{National Center for Supercomputing Applications, University of Illinois at Urbana-Champaign, Urbana, IL 61801, USA}

\begin{abstract}
Remote entanglement between widely separated qubits is a fundamental quantum phenomenon and a critical resource for quantum information applications. 
Generating entanglement between independent qubits separated by arbitrary, potentially large distances requires propagating quantum states, and is typically achieved using pulsed protocols combining distinct steps of local entanglement generation followed by distribution. 
This necessity raises an intriguing question: 
Can remote entanglement be stabilized indefinitely, instead of only periodically regenerated and redistributed after decay?
Here, we demonstrate that this is indeed possible, reporting autonomous stabilization of entanglement between two separate superconducting-qubit devices.
Combining nonreciprocal waveguide coupling and local driving, we experimentally realize a symmetry-based coherent quantum-absorber scheme in a cascaded network. 
We quantify the degree of entanglement through quantum state tomography, finding that the protocol’s entangling power is severely limited by imperfections that break the required symmetry.
We show, however, that a modified protocol based on an alternate symmetry is far more robust, enabling us to achieve a concurrence approaching 0.5, a limit set only by local loss in the network. Our results enable on-demand delivery of high-fidelity entanglement in modular quantum processors and networks and pave the way for autonomously protecting distributed quantum information.
\end{abstract}
\maketitle

Remote entanglement describes quantum correlations between macroscopically separated particles, illustrating strikingly the tension between quantum and classical physics. 
While this `spooky action at a distance' was originally devised to point out difficulties in interpreting quantum mechanics~\cite{einstein_can_1935}, remote entanglement has in recent decades emerged as a cornerstone of quantum information applications~\cite{ekert_quantum_1991,bennett_teleporting_1993,gottesman_demonstrating_1999} and is nowadays generated routinely in many experimental platforms~\cite{julsgaard_experimental_2001,chou_measurementinduced_2005,moehring_entanglement_2007,bernien_heralded_2013,hofmann_heralded_2012,roch_observation_2014a,riedinger_remote_2018}.
Remarkably, it is possible to observe entanglement correlations between qubits that are separated by arbitrary distances and not explicitly interacting with one another; 
this behavior underlies both the fundamental appeal of entanglement and its practical utility.
While it is possible to realize such a `nonlocal' entanglement, the origin of its generation lies in local interactions.
Remote entanglement can be realized by first generating entanglement between qubits locally, and then subsequently separating them. 
In practice, this concept is combined with projective measurements~\cite{duan_longdistance_2001} or controlled  absorption~\cite{cirac_quantum_1997a} to prepare remote entanglement in a stepwise fashion.

It is intriguing to ask whether remote entanglement can be preserved in the steady state, i.e., stabilized indefinitely without re-preparing it following its decoherence.
The fundamental question is whether there is a mechanism for generating entanglement between independent qubits at arbitrary distances without the need for repeated preparation cycles consisting of distinct steps of local entanglement generation followed by distribution.
Additionally, the achievement of `always-on' remote entanglement has important practical implications.
Remote entanglement is a critical resource in quantum networks and measurement-based quantum computation that is consumed on-demand~\cite{humphreys_deterministic_2018,chouDeterministicTeleportationQuantum2018}.
Steady-state entanglement would avoid delays in its availability and obviate the need for its active re-preparation after it inevitably decays due to decoherence.

\begin{figure*}[t]
	\centering
	\includegraphics{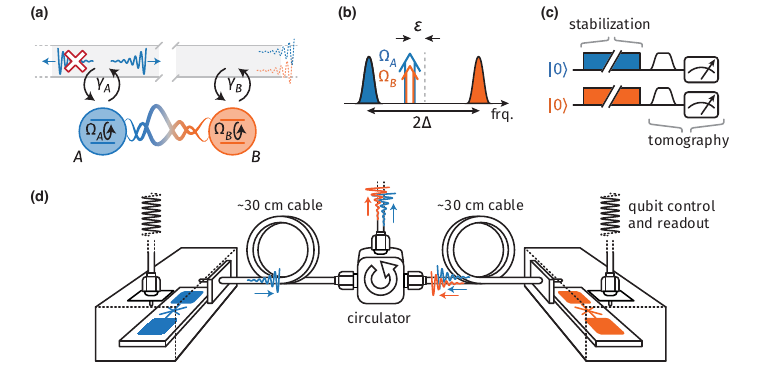}
	\caption{
        \textbf{Setup for driven-dissipative remote entanglement stabilization.}
        (\textbf{a}) Two qubits coupled to a unidirectional waveguide can be driven into an entangled steady state.
        Without loss, this coherent quantum-absorber system can exhibit an entangled dark state.
        (\textbf{b, c}) To demonstrate this idea, the two qubits are driven into the steady state at the same drive frequency, at a detuning $\varepsilon$ from their center frequency. 
        Two-qubit quantum state tomography reveals the entanglement.
        (\textbf{d}) The experimental setup consists of two independent superconducting transmon qubit devices, connected by a low-loss coaxial cable link that is made unidirectional through a microwave circulator.
        Qubit state readout is performed through a dispersively coupled readout resonator that is not shown in this schematic (Appendix~\ref{app:setup}).
		}
	\label{fig:concept}
\end{figure*}

A natural approach for realizing steady-state entanglement is presented by bath engineering, a paradigm for autonomously stabilizing arbitrary quantum states.
The key ingredient here is the realization of collective dissipation and continuous driving such that the target state becomes the non-equilibrium steady state~\cite{poyatos_quantum_1996}.
This idea has already been used to stabilize entanglement locally:
In single quantum devices, the inter-qubit distance can be matched to interaction-mediating waves in oscillators or waveguides acting as a suitable bath that is shared between qubit pairs~\cite{lin_dissipative_2013,shankar_autonomously_2013,kimchi-schwartz_stabilizing_2016,brown_trade_2022a,shah_stabilizing_2024}.
This approach, however, fundamentally treats qubits and mediating modes as a joint system, where locations and parameters of components are fine-tuned to each other.
In contrast, steady-state entanglement between separated qubits in a network requires the realization of a nonlocal bath that makes no assumptions about the inter-qubit distance.

Previous theoretical work has proposed approaches for engineering two-qubit dissipation that can stabilize distance-independent entanglement.
A suitable bath for this task can be realized by distributing two-mode squeezed vacuum~\cite{kraus_discrete_2004} or coupling qubits radiatively to a unidirectional waveguide~\cite{krauter_entanglement_2011,stannigel_drivendissipative_2012a,pichler_quantum_2015a}.
An experimental demonstration of these schemes has, however, remained elusive.
The crucial difficulty for implementing them is the precise tailoring of the required two-qubit dissipation:
First, the correlated, nonlocal dissipation must strongly dominate over any local loss, which requires an efficient distribution of correlated noise.
Second, loss and excitation must be matched correctly for both qubits in order to stabilize an entangled target state.
Because our goal is the entanglement of independent devices, these challenges additionally raise the broader question of how to realize the desired correlated dissipation without requiring fine-tuned symmetries between the remote qubits.

Here, we report the experimental demonstration of steady-state remote entanglement between independent superconducting qubit devices.
We have realized a low-loss cascaded quantum network, with the qubits separated by approx.\ 60\,cm of coaxial cable and a circulator that makes the transmission channel unidirectional.
We combine the effective chiral waveguide coupling with local drives, allowing us to demonstrate driven-dissipative entanglement using a so-called coherent quantum absorber (CQA) scheme~\cite{stannigel_drivendissipative_2012a,pichler_quantum_2015a}.
We find that inevitable imperfections violate symmetry assumptions in the theoretical construction of the protocol, resulting in an amount of entanglement that falls short of what can be explained by the loss rates of the network.
This discrepancy is rooted in the fact that the CQA scheme requires a spatial symmetry between the remote parties that cannot be met with sufficient precision in a real experiment.
We overcome this challenge by identifying a connection to a protocol that mimics two-mode squeezing and does not require such a symmetry.
We find that we can tailor drive parameters \emph{in situ} to implement this approach and dramatically increase the amount of stabilized entanglement stabilized.
We observe an optimal concurrence of $\conc \approx 0.5$, showing a compelling path toward autonomous stabilization of high-fidelity entanglement in quantum networks.

\begin{figure*}[t]
        \centering
        \includegraphics{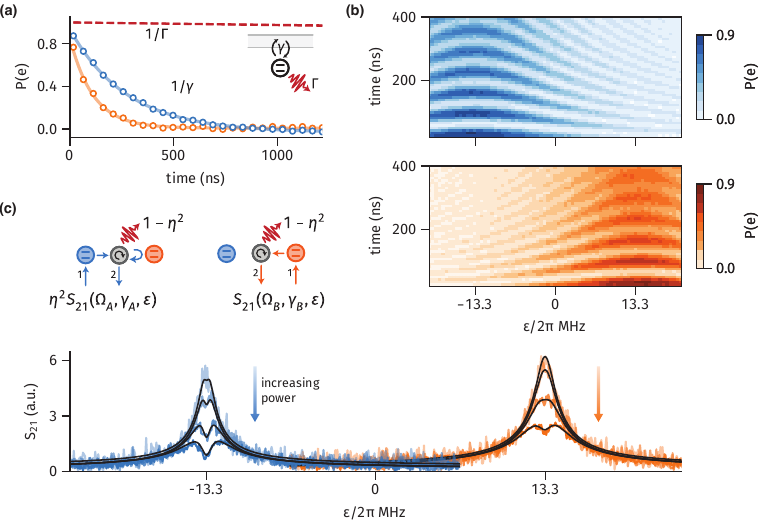}
        \caption{
            \textbf{Network characterization.}
            (\textbf{a}) The decay time of each qubit from the excited ($\ket{1}$) to the ground ($\ket{0}$) state is determined by the waveguide coupling, and measured in a conventional $T_1$ experiment.
            Symbols: experimental data. 
            Solid lines: exponential fits, yielding $T_1$ times of 300\,ns for qubit A (blue), and 130\,ns for qubit B (orange).
            Dashed lines: for comparison, qubit relaxation measured prior to making the waveguide connection.
            (\textbf{b}) Simultaneously driven Rabi oscillations on both qubits.
            The qubits are $26.6$ MHz detuned from each other.
            The CQA model predicts entanglement stabilization at $\varepsilon=0$.
            (\textbf{c}) Measurement of transmission loss through coherent scattering.
            The scattering responses ($|S_{21}|$) of qubits A and B corresponding to increasing input powers are plotted in solid blue and orange lines.
            Black lines: simultaneous fit over all powers, and both qubits, yielding a transmission efficiency $\eta^2 = 0.96 \pm 0.01$.
            See \cref{app:Network Characterization} for details.
            }
        \label{fig:characterization}
    \end{figure*}

\subsection*{Coherent quantum absorber scheme}

Our experiment closely follows the originally proposed CQA protocol~\cite{stannigel_drivendissipative_2012a,pichler_quantum_2015a}, illustrated in Fig.~\ref{fig:concept}a:
Two qubits, Alice (A) and Bob (B), are coupled radiatively to a unidirectional waveguide with rates $\gamma_A$ and $\gamma_B$, where we first assume that $\gamma \equiv \gamma_A = \gamma_B$.
This situation corresponds to a cascaded quantum system, where the emission of the upstream qubit (A) drives the downstream qubit (B)~\cite{gardiner_driving_1993a,carmichael_quantum_1993a, gunin_quantum_2023, dmitriev_direct_2025}.
In this setup, under the assumption of no local dissipation, applying local Rabi drives to both qubits is predicted to produce a dark state of the form $\ket{\psi} \propto \ket{00} + \alpha(\ket{01} - \ket{10})$, where $\alpha = \sqrt{2}\Omega / (2\Delta-i\gamma)$.
Here, $\Omega$ is the common Rabi frequency for both qubits and $2\Delta$ is the detuning between the qubits, with the drives applied at the center frequency ($\varepsilon = 0$) (Fig.~\ref{fig:concept}b,c).
The emergence of this state can be understood as a destructive interference of emission of the two qubits, with no photons leaving the system.
Thus, simply applying appropriately matched drives to the two qubits yields the maximally entangled singlet state, $\ket{S} \equiv (\ket{01} - \ket{10})/\sqrt{2}$ in the limit of large $\Omega$. 
This scheme fulfills a strict definition of remote entanglement:
The two qubits are `disconnected', and the unidirectional waveguide coupling makes the protocol fully distance-independent~\cite{pichler_quantum_2015a}.

The experiment faces several crucial challenges:
First, it requires a setup with strong spatial symmetries in the form of matched $\gamma$ and $\Omega$; this poses a difficulty for the construction of a network of independent devices.
Second, inevitable decoherence in the form of qubit relaxation, dephasing, and waveguide loss dramatically limit performance.
These loss rates compete with the characteristic relaxation rate of the system, resulting in several important consequences~\cite{irfan_loss_2024a,brown_trade_2022a,pocklington_accelerating_2025}.
Local and uncorrelated loss channels impose a ceiling on the maximum entanglement that can be stabilized, and they shrink the parameter space in which high degrees of entanglement can be found.
While it is possible to overwhelm these loss rates with correlated dissipation through an increase in $\gamma$, this strategy makes entanglement detection challenging: 
After turning off the stabilization drives, the state immediately decays with a rate $\gamma$. 
A practical limit on $\gamma$ is thus imposed by how fast readout can be performed.
These considerations demand the suppression of qubit and waveguide loss to the greatest extent possible.

\begin{figure*}[t]
       \centering
       \includegraphics{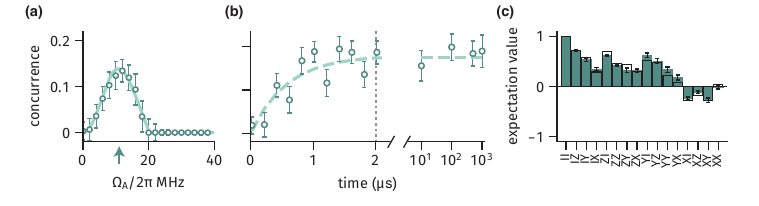}
       \caption{
           \textbf{Steady-state entanglement with the CQA scheme.} 
           (\textbf{a}) Measured concurrence (markers) as a function of Rabi drive amplitude, with $\Omega_A = \Omega_B$ and $\varepsilon=0$.
           Solid line: Fit to a master equation simulation that includes independently measured loss parameters as well as drive-induced frequency shifts of the transmons (Appendix~\ref{app:analysis}).
           (\textbf{b}) Measured concurrence as a function of drive duration for $\Omega/2\pi = 12\,\mathrm{MHz}$, indicated by the green arrow in (a). The dashed gray line indicates the drive duration that is used to prepare the steady state for each $\Omega_A$ in (a).
           The discrepancy between the average concurrence value in the plateau of Fig.~\ref{fig:CQA}b and the peak concurrence in Fig.~\ref{fig:CQA}a is explained in Appendix~\ref{app:sensitivity}.
           (\textbf{c}) Measured expectation values (solid bars) of all two-qubit correlators for the steady state with $\Omega/2\pi = 12\,\mathrm{MHz}$.
           Black outlines: simulated expectation values.
           }
       \label{fig:CQA}
   \end{figure*}

\subsection*{Experimental setup}

Our experimental setup addresses the above considerations as depicted schematically in Fig.~\ref{fig:concept}d.
Superconducting transmon qubits can show highly efficient and controllable qubit-waveguide coupling~\cite{astafiev_resonance_2010, vanloo_photonmediated_2013}, making them an ideal platform for testing the CQA scheme.
To realize a low-loss cascaded network we employ fixed-frequency transmons housed in low-loss waveguide enclosures~\cite{axlineArchitectureIntegratingPlanar2016a}.

We connect them with a low-loss unidirectional waveguide realized by combining custom microwave connections that have recently enabled efficient inter-device gates~\cite{mollenhauer_highefficiency_2025} with a low-loss microwave circulator.
In this setup, we expect significantly lower loss than in previously demonstrated cascaded systems used for quantum state transfer, where total loss rates were on the order of 25\,\%~\cite{axline_ondemand_2018b,campagne-ibarcq_deterministic_2018b,kurpiers_deterministic_2018}.

We characterized the qubit-waveguide coupling by measuring the qubit relaxation rates in a $T_1$ experiment (Fig.~\ref{fig:characterization}a).
Prior to making the waveguide connection, both qubits had relaxation times of approximately \qty{40}{\micro\second}; we can thus safely assume that $\gamma = 1/T_1$.
From these measurements we infer $\gamma_A/2\pi = 0.53\,\mathrm{MHz}$ and $\gamma_B/2\pi = 1.22\,\mathrm{MHz}$, close to the target value of 1\,MHz and limited in precision by the mechanical assembly of qubit-waveguide coupling~\cite{mollenhauer_highefficiency_2025}.
We note that this inevitable mismatch breaks the strict exchange symmetry between the two qubits that is at the heart of the CQA entangling mechanism.
As we will see, however, this mismatch does not preclude entanglement stabilization and we are able to overcome limitations imposed by the mismatch with an enhanced protocol.
The qubit resonant frequencies are $2\Delta/2\pi = 26.6\,\mathrm{MHz}$ apart.
Entanglement stabilization is thus expected when driving both qubits at $\omega_d/2\pi = 4.665\,\mathrm{GHz}$, corresponding to $\varepsilon = 0$ (Fig.~\ref{fig:characterization}b).
All system parameters are detailed in Appendix~\ref{app:setup}.


Before moving on to the detection of entanglement, we performed a transmission experiment (Fig.~\ref{fig:characterization}c) to independently confirm that the directional waveguide has low loss. 
Here, we effectively compare the magnitude of two separately acquired scattering signals off the two qubits.
The key is that the scattered signal from qubit A traverses the same path as that from qubit B, plus the unidirectional link.
The shape of the scattered signal is determined by the drive amplitude, $\Omega$, and the qubit-waveguide coupling, $\gamma$. 
The relative magnitude of the emitted signals, normalized by the drive amplitude, directly encodes the transmission efficiency $\eta^2$. 
By simultaneously fitting the scattering responses of both qubits over multiple powers to a numerical model, we extract $\eta^2 = 0.96 \pm 0.01$. 
This measurement confirms the realization of a low-loss cascaded network and agrees with the predicted reduction in loss achieved through the use of custom microwave connections.
More details on this analysis are given in Appendix~\ref{app:Network Characterization}.

\subsection*{Stabilization and detection of entanglement}
\label{Stabilization and detection of entanglement}
With the setup characterized, the protocol for stabilizing and detecting entanglement is conceptually simple (Fig.~\ref{fig:concept}b,c):
We apply drives on both qubits simultaneously, at the same drive frequency $\omega_d$ and  with Rabi rates $\Omega_{A,B}$.
After the qubits have reached their steady state (approximately \qty{1}{\micro\second}, see below) we turn off the stabilization drives and immediately perform full quantum state tomography.
Tomography is performed by single-qubit pre-rotations setting the measurement axis, followed by a standard dispersive qubit measurement of the $\sigma_z$ component of the qubit state~\cite{blaisCavityQuantumElectrodynamics2004a}.
By measuring a complete set of measurement operators in this way, the full quantum state of the system can be reconstructed.

A crucial challenge in accurately inferring the quantum state of the qubits is their short lifetime.
The waveguide coupling leads to $T_1$ times that are non-negligible compared to the duration of the pre-rotations, and are on the order of the fastest reported dispersive qubit measurements~\cite{walterRapidHighFidelitySingleShot2017a,sunadaFastReadoutReset2022b}. 
As a consequence of these timescales, high-fidelity single-shot readout is infeasible and the measurement operators do not correspond to simple Pauli operators.
To overcome this challenge, we combine numerical simulations of the relaxation process with quantum detector tomography.
This procedure characterizes the measurement in the form of positive operator-valued measures (POVMs), and their knowledge lets us reconstruct the quantum state using a convex optimization~\cite{lundeen_tomography_2009}.
In this way, we reliably reconstruct the two-qubit density matrix, $\rho$, with small uncertainties (Appendix~\ref{app:analysis}).
We then determine directly the degree of observed entanglement using the concurrence, an entanglement monotone, where $\conc > 0$ is proof of distillable entanglement and $\conc = 1$ corresponds to maximally entangled states.

\begin{figure*}[t]
	\centering
	\includegraphics{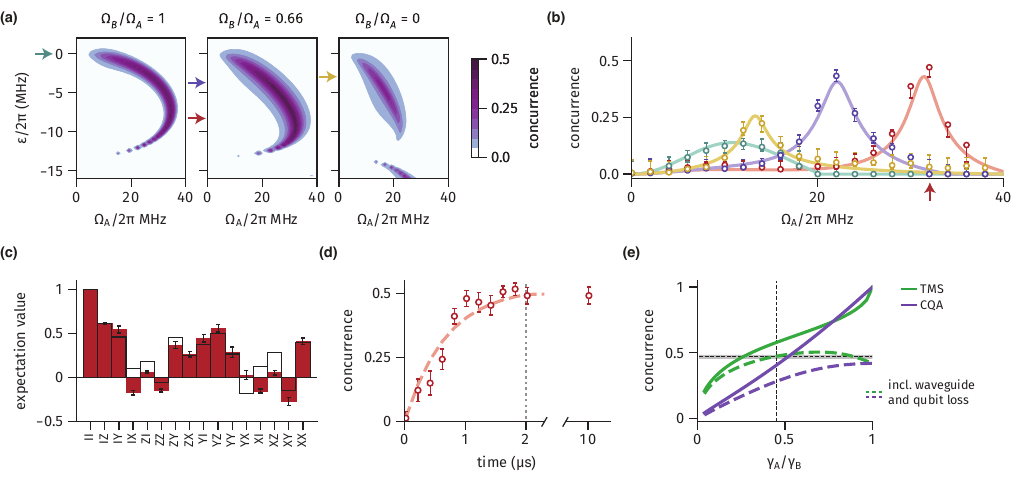}
	\caption{
        \textbf{Improving concurrence with synthetic squeezing symmetry.}
        (\textbf{a}) Simulated steady-state concurrence as a function of drive detuning $\varepsilon$ and drive strength $\Omega_A$ for different ratios $\Omega_B/\Omega_A$ (see Appendix~\ref{app:modeling} for details).
        (\textbf{b}) Measured steady-state concurrence (markers) as a function of drive strength $\Omega_A$ for parameters indicated by colored arrows in (a). 
        Solid lines are fits to master equation simulations. 
        The teal curve was shown in Fig.~\ref{fig:CQA}a.
        (\textbf{c}) Two-qubit correlators for the state with the highest concurrence (maximum concurrence of the red curve in (b)). 
        The measured state has $96\%$ fidelity with the simulated state.
        (\textbf{d}) Measured concurrence as a function of drive duration for parameters corresponding to the highest concurrence in (b), indicated by the red arrow. The dashed gray line indicates the drive duration that is used to prepare the steady state for each $\Omega_A$ in (b).
        (\textbf{e}) Simulated maximum concurrence achievable in the present system as a function of the ratio of qubit-cable coupling strengths.
        The purple (green) lines show the maximum concurrence that can be stabilized using the CQA (synthetic TMS) symmetry.
        Dashed lines: optimal performance given the system's intrinsic loss rates (transmission loss, dephasing and intrinsic relaxation rates); solid lines: predicted performance in the absence of any loss.
        Black dashed lines: the experimental system investigated in this work; gray shading indicates uncertainty in the concurrence.
		}
	\label{fig:landscape}
\end{figure*}


We first investigated the CQA prescription by driving at $\varepsilon = 0$ and with both Rabi rates equal, $\Omega_A = \Omega_B \equiv \Omega$.
We expect, in general, no entanglement at small $\Omega$; and in the presence of finite local losses we expect no entanglement at very large $\Omega$. 
The reason for the latter is that the relaxation into the singlet state slows down, and local losses dominate over entanglement stabilization; this effect is due to a shrinking of the dissipative gap~\cite{irfan_loss_2024a}.
There is thus a local maximum in $\conc$ at a finite $\Omega$, with a location depending on specific system parameters.
To verify this expectation, we swept $\Omega$ and for each point determined $\conc$~(Fig.~\ref{fig:CQA}a).
Indeed, we find clear evidence of remote entanglement, with a maximal $\conc = 0.135^{+0.021}_{-0.024}$ occurring at $\Omega/2\pi = 12\,\mathrm{MHz}$.
To verify that this is indeed the steady state, we varied the length of the drive pulse; 
after a short stabilization time of approx.\ \qty{1}{\micro\second}, the concurrence remains stable up to 1\,ms, the longest time we have tested (Fig.~\ref{fig:CQA}b).
Based on this stabilization time, all steady states reported in this paper are prepared with \qty{2}{\micro\second} drives, unless noted otherwise.

The fact that the state is stable over more than three orders of magnitude beyond the longest lifetime of any part of our system is strong evidence of its steady-state nature.
To confirm our theoretical understanding of these results, we have performed numerical simulations of this system and found excellent agreement not only with the evolution of entanglement as a function of $\Omega$, but also the specific state generated:
The reconstructed steady state with the highest concurrence has a fidelity of 0.99 with the predicted one (Fig.~\ref{fig:CQA}c).

This result is a remarkable confirmation of earlier theoretical predictions, and a central result of this study.
The measured concurrence, however, is significantly lower than the bound of 0.5 set by local decoherence rates (qubit relaxation, qubit dephasing, and waveguide loss) and the qubit-waveguide coupling mismatch.
Part of this discrepancy can be attributed to drive-induced qubit frequency shifts moving the detuning away from the optimal operating point (Appendix~\ref{app:sensitivity}).
However, the best possible performance of the CQA protocol (predicted to be $\conc = 0.29$) still falls significantly short of the limit set by local loss (Appendix~\ref{app: Expectations and best performance}).

This raises the questions (1) how to physically interpret the lower degree of inferred entanglement, and (2) whether there are strategies that allow the restoration of the optimally achievable (i.e., loss-bounded) amount of entanglement.

\subsection*{Symmetry conditions for optimal entanglement}

The optimal concurrence is ultimately limited by the mismatch in the qubit-waveguide coupling strengths, which breaks the symmetry required for perfect absorption.
Theoretical work has shown that the CQA scheme relies on a `hidden time reversal symmetry' that underpins the existence of a pure entangled steady state in the ideal case~\cite{robertsHiddenTimeReversalSymmetry2021}.
This symmetry is not robust to qubit-waveguide coupling mismatch and therefore the mismatch limits the degree of entanglement that can be stabilized.
Because some degree of disorder is inevitable in a quantum network of independent devices, this raises an important fundamental question: 
Is there an alternative symmetry that can be exploited to achieve a higher degree of entanglement if the symmetry assumptions of CQA are not met exactly?

An alternative symmetry that leads to pure steady-state entanglement is appropriately correlated excitation and relaxation, which can be achieved by injecting two-mode squeezed (TMS) vacuum~\cite{kraus_discrete_2004}.
TMS is, however, not required; it is sufficient to realize the essential symmetry in the dissipators, an approach that has been referred to as `synthetic squeezing'~\cite{govia_stabilizing_2022}.
Interestingly, this symmetry can be enforced even in the presence of arbitrarily large qubit-waveguide coupling mismatches.
By transforming our system into an appropriate frame, we find that this can be achieved here when drive strengths obey the relation $\gamma_A \Omega_A^2 = \gamma_B \Omega_B^2$.
Further, the effective Rabi rates must be matched by fulfilling $\varepsilon = (\Omega_B^2 - \Omega_A^2)/4\Delta$. We can interpret these two conditions as the stimulated-emitted photons from the two qubits having the same amplitude and frequency.
A detailed derivation for this optimal drive condition is presented in Appendix~\ref{app:TMS}.

This analytical prediction is easily confirmed by numerical simulations (Fig.~\ref{fig:landscape}a).
These predict that changes in the ratio of the drive strengths, $\Omega_B/\Omega_A$, have a pronounced effect on the parameter range in which entanglement can be found, and on the maximally attainable entanglement.
At precisely the drive ratio at which the new symmetry condition is fulfilled for our system, $\Omega_B/\Omega_A = 0.66$, simulations predict the largest concurrence.
We have experimentally verified these predictions by representative sweeps of $\Omega_A$ for a set of $\varepsilon$ and $\Omega_B/\Omega_A$, finding excellent agreement with the numerical model and confirming that the highest degree of entanglement is observed when our modified symmetry condition is met (Fig.~\ref{fig:landscape}b).
The states are again in very good agreement with numerical simulations (Fig.~\ref{fig:landscape}c) and are confirmed to be steady states of the two-qubit system (Fig.~\ref{fig:landscape}d).
With the new symmetry condition applied, we detect a maximal concurrence of $\conc = 0.471^{+0.004}_{-0.041}$.
Importantly, this value is in agreement with our original expectation of the maximal concurrence achievable, given the loss in the system.
Indeed, a numerical analysis shows that this concurrence is optimal in the sense that it meets what could have been achieved with CQA in the best case, i.e., if $\gamma_A$ and $\gamma_B$ had been matched (Fig.~\ref{fig:landscape}e).

Finally, we note that the system stabilizes entanglement even when the downstream drive is turned off, $\Omega_B = 0$ (Fig.~\ref{fig:landscape}a,b).
While this observation appears counterintuitive, it can be understood as a consequence of the nonreciprocal waveguide coupling that results in correlated dissipators:
A single drive applied on qubit A generates population in the state $\ket{10}$, which can be written as a linear combination of singlet and triplet entangled states.
Because the correlated relaxation results in a preferred decay of the triplet state~\cite{pichler_quantum_2015a}, we are left with preferential population in the singlet, and thus a net-entangled state.
A more detailed analysis of this case is presented in Appendix~\ref{app: Upstream Drive}

\subsection*{Summary and Outlook}

We have demonstrated steady-state remote entanglement, between spatially separated and fully independent qubit devices that do not interact coherently.
The performance of the originally proposed experimental configuration is limited by inevitable disorder.
Remarkably, the construction of a synthetic squeezing symmetry allows us to achieve the maximal amount of entanglement allowed by local losses.
It is an intriguing question whether similar avenues can be used to enhance schemes that utilize distributed two-mode squeezed vacuum~\cite{andres-juanes_entangling_2025}.
Capitalizing on this enhanced scheme, we are able to observe an amount of remote entanglement that is competitive with deterministic, pulsed remote entanglement generation in superconducting qubit networks~\cite{axline_ondemand_2018b,campagne-ibarcq_deterministic_2018b,kurpiers_deterministic_2018}, and on par with reports of entanglement stabilization within single superconducting devices~\cite{shankar_autonomously_2013,kimchi-schwartz_stabilizing_2016,brown_trade_2022a,shah_stabilizing_2024}.

It is an interesting question how steady-state remote entanglement generation can be compared to traditional protocols~\cite{duan_longdistance_2001,cirac_quantum_1997a}.
A key difference is the suspectibility to waveguide loss: 
In `stroboscopic' approaches, photon loss directly results in a reduction of the attainable degree of entanglement. 
This is not the case for the stabilization approach shown here, where loss can always be overcome by stronger waveguide coupling, $\gamma$.
This advantage comes at a cost, however: 
Stronger waveguide coupling results in a faster decay of the entangled state once stabilization drives are turned off, and higher-fidelity entangled states may take a longer time to stabilize.
Overcoming these challenges requires additional complexity in experiments~\cite{irfan_loss_2024a,brown_trade_2022a}.
At this time we thus conclude that the best choice of protocol for entanglement delivery is likely dependent on the details of the specific experimental platform.

Using the demonstrated protocol in practice for entanglement delivery is conceptually straight-forward by supplementing the waveguide-coupled qubits with additional qubits to which the entanglement can be transferred rapidly on-demand or in the steady state~\cite{irfan_loss_2024a,lingenfelter_exact_2024}.
In this way, the entanglement can be stored even after the stabilization drives are turned off.
Further improvements to the degree of attainable entanglement be achieved by either increasing the waveguide coupling or by lowering transmission loss through integration with lossless chirality~\cite{kannan2023demand,joshiResonanceFluorescenceChiral2023a,caoParametricallyControlledChiral2024}.

Our work further paves the way for autonomous stabilization and protection of multi-qubit, distributed entanglement.
Theoretical work has shown that our experimental setup can be extended to stabilize multiple entangled pairs simultaneously~\cite{lingenfelter_exact_2024} and that such a system has remarkable, built-in loss resilience~\cite{irfan_loss_2024a}. 
In combination with proposed approaches for autonomous entanglement distillation~\cite{vollbrechtEntanglementDistillationDissipation2011} or quantum error correction~\cite{kerckhoff_designing_2010}, our work can thus enable the autonomous protection of entanglement in extended quantum device networks.
This perspective provides an avenue for investigating distributed, non-equilibrium many-body states and for on-demand delivery of high-fidelity entanglement in modular quantum processors.

\section*{Acknowledgements}
The research was carried out in part in the Materials Research Lab Central Facilities and the Holonyak Micro and Nanotechnology Lab at the University of Illinois.
We thank K. Chow and R. Goncalves for help with qubit device fabrication.
We thank M.~Hatridge for valuable discussions on optimizing qubit readout.
We thank A.~Kou for valuable feedback on the manuscript.

\paragraph*{Funding:}
This work was supported by the National Science Foundation under Award No.~2016136 (A.I., M.Y., A.L., X.C., A.C., W.P.); 
by the Air Force Office for Scientific Research under Grant No.~FA9550-24-1-0354 (A.I., K.S., M.Y., A.L., A.C., W.P.); 
by the Army Research Office under Grant No.~W911NF-23-1-0077 (M.Y., A.L., A.C.);
and by IBM through the IBM-Illinois Discovery Accelerator Institute (M.M., X.C., W.P.).
A.C.\ acknowledges support from the Simons Foundation through a Simons Investigator Award (Grant No.~669487).

\paragraph*{Author contributions:}
A.I.\ designed and fabricated the device. 
A.I., K.S., and M.M.\ built the setup and acquired the experimental data.
A.I. and K.S.\ analyzed the data.
M.Y., A.I., K.S., and X.C.\ developed numerical simulations. 
M.Y.\ and A.L.\ developed the analytical framework for optimal drive conditions.
A.I., K.S., M.Y., and W.P.\ wrote the manuscript, with contributions from all authors.
A.C. and W.P.\ supervised the project.

\appendix
\section{Experimental Setup}
\label{app:setup}
\begin{figure*}[t]
    \centering
    \includegraphics{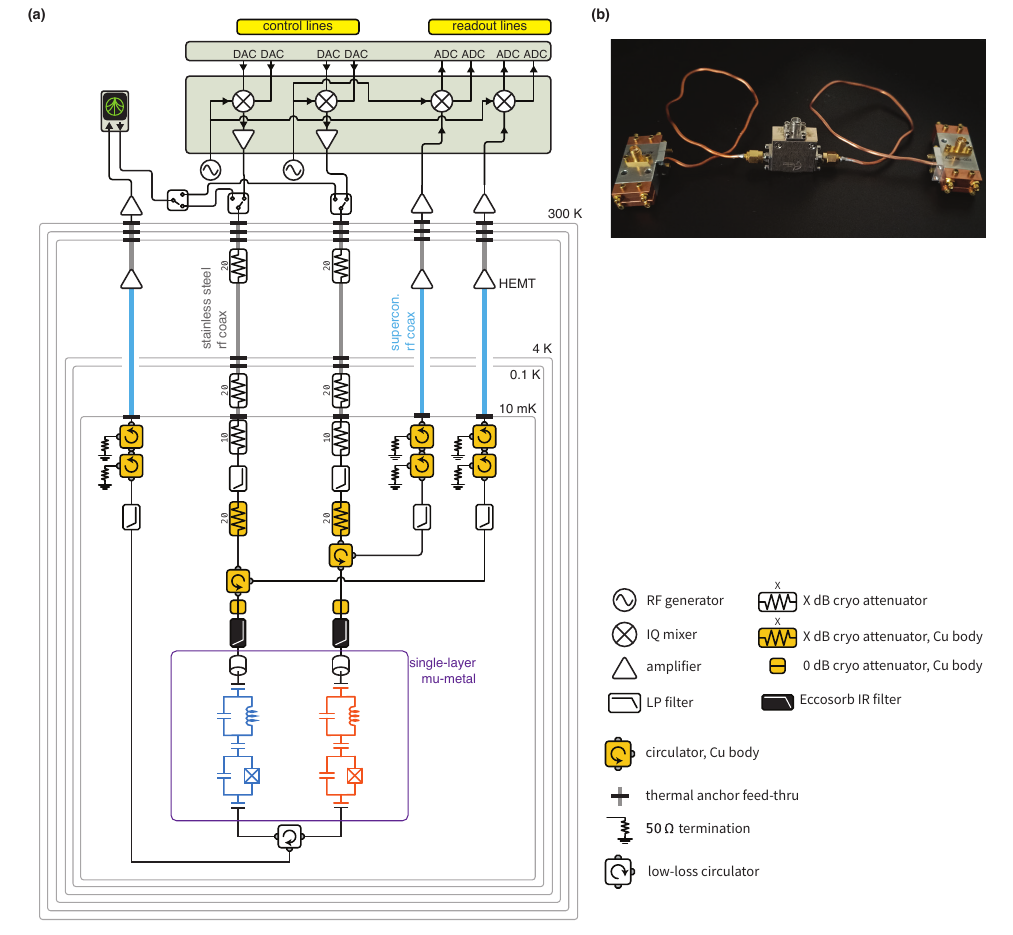}
    \caption{
    \textbf{Experimental Setup}.
    (\textbf{a}) Most of the wiring diagram follows standard best practices~\cite{krinner2019engineering}.
    Noteworthy components: 
    \textbf{low-loss circulator}: Quinstar Technology QCY-G0400801AUZ 4-8 GHz cryogenic circulator with $0.15$ dB typical insertion loss and $< 0.08$ dB insertion loss at $6$ GHz.
    \textbf{Probe Line} Low pass filters: 12 GHz (K$\&$L 5L250-10200).
    \textbf{Output line}: double-stage isolators (Low Noise Factory LNF-ISCIC4$\_$12A); HEMT amplifier (Low Noise Factory LNF-LNC4$\_$8C); room temperature low-noise amplifier (Low Noise Factory LNF-LNR4$\_$14C).
    \textbf{Room temperature electronics}: signal modulation and demodulation: Quantum Machines OPX+ and Quantum Machines Octave; VNA: Keysight P9374A.
    (\textbf{b}) The assembly of the experiment.
    }
    \label{suppfig:Assembly and wiring}
\end{figure*}

\paragraph{Cryogenic measurement setup} Devices were cooled to 10 mK in an Oxford Instruments Triton 500 dilution refrigerator. 
A schematic of setup and device wiring is shown in Fig.~\ref{suppfig:Assembly and wiring}a.

\paragraph{Qubit Devices} Each device consists of a fixed-frequency transmon coupled to a stripline resonator (architecture described in ~\cite{axlineArchitectureIntegratingPlanar2016a}).
Fabrication proceeds as follows: a 200 nm NbTiN film is sputtered onto a sapphire wafer. 
The transmon capacitor pads and the stripline resonator are defined by photolithography, followed by plasma etching. 
Josephson junctions are formed via the Dolan bridge technique using double-angle Aluminum evaporation $\pm40^\circ$ at 0.3 nm/s. Prior to Aluminum deposition, the NbTiN native oxide is removed by Argon milling to allow contact between the junction and the pads.
These devices are placed inside sub-cutoff waveguide enclosures made of 6061 aluminum alloy.
The measured device parameters can be found in Table~\ref{tab:transmon parameters}.

\paragraph{Network Assembly} SMA connectors were soldered to both ends of an approximately 60 cm long copper coaxial cable, followed by verification of low return loss using a vector network analyzer.
The cable assembly was then divided into two parts by cutting it through the center.
The cut end of each coaxial cable was clamped to a transmon module using the mechanism described in \cite{mollenhauer_highefficiency_2025}, while the other end was attached to the circulator.

\begin{table} 
	\centering
	\caption{
    \textbf{System parameters.}
    Measured qubit and readout resonator parameters.
    The intrinsic relaxation times of the qubits were measured in a separate experiment where the qubits were not coupled to the waveguide.
    }
	\label{tab:transmon parameters} 

	\begin{tabular}{lccr} 
		\\
		\hline\hline
		Qubit A &  & Value\\
		\hline
		Qubit frequency & $\omega_A / 2\pi$ & 4.6522 GHz\\
        Resonator frequency & $\omega_{RA} / 2\pi$ & 7.5089 GHz\\
		  Relaxation time & $T_1$ & 302(2) ns\\
		Ramsey decay time & $T_{2R}$ &  476(6) ns\\
        Intrinsic relaxation time & $T_1$ & \qty{38}{\micro\second}\\
		\hline
        Qubit B &  & Value\\
		\hline
		Qubit frequency & $\omega_B / 2\pi$ & 4.6787 GHz\\
        Resonator frequency & $\omega_{RB} / 2\pi$ & 7.5445 GHz\\
		  Relaxation time & $T_1$ & 133(2) ns\\
		Ramsey decay time & $T_{2R}$ & 266(7) ns\\
        Intrinsic relaxation time & $T_1$ & \qty{44}{\micro\second}\\
		\hline\hline
	\end{tabular}
\end{table} 


\section{Measuring transmission loss}
\label{app:Network Characterization}
\paragraph{Expected error budget} Previously demonstrated cascaded quantum network experiments have estimated around $15\,\%$ transmission loss \cite{axline_ondemand_2018b, campagne-ibarcq_deterministic_2018b, kurpiers_deterministic_2018}.
By eliminating two SMA connectors (0.06dB each) and using a low-loss circulator (specified as 0.15~dB loss, see Fig.~\ref{suppfig:Assembly and wiring} for model number), we expect the transmission loss in our cascaded network to be around $5\%$.

\paragraph{Asymmetric scattering measurement} Using a vector network analyzer (VNA), we measure the transmission spectrum ($|S_{21}|$) for each qubit.
This is achieved as follows.
The drive from the VNA (port 1) enters the drive port of the qubit module, scatters off the qubit, and is emitted into the waveguide. 
The output signal leaves the network from the third port of the circulator before entering the VNA (port 2).
The path of the scattered signal for each qubit is illustrated in Fig.~\ref{fig:characterization}c.
As the input power is increased, the qubit saturates, producing a line shape that departs from the Lorentzian response of a harmonic oscillator.
The shape of the spectrum is determined by the qubit-waveguide coupling $\gamma$ and the drive amplitude $\Omega$.
Note that for qubit A, the scattered signal traverses the circulator twice (compared to once for qubit B).
Therefore, the relative magnitude of the emitted signals, normalized by the drive amplitude, directly encodes the transmission efficiency $\eta^2$.

We model this measurement by simulating a driven transmon using the following Hamiltonian:
\begin{align}
    \hat{H} = \Delta \hat{a}^{\dagger}\hat{a} &+ \alpha (\hat{a}^{\dagger}\hat{a}^{\dagger}\hat{a}\hat{a}) + \Omega (\hat{a}^{\dagger} + \hat{a})
\end{align}
where $\Delta$ is the drive detuning from the transmon frequency and $\alpha$ is the anharmonicity of the transmon.
The qubit is weakly coupled to the port through which it is driven and strongly coupled to the coaxial cable through which it decays.
Therefore, the emitted field can be described by $\hat{a}_{out} = \sqrt{\gamma} \hat{a}$, where $\gamma$ is the qubit-waveguide coupling strength known by measuring the relaxation time of the qubit.
The magnitude of the spectrum measured at the third port of the circulator can be modeled as the expectation value of $\hat{a}_{out}$, normalized by $\Omega$. 
Since we know that the scattered signal from qubit A traverses the circulator twice, we multiply its spectrum by $\eta^2$.
The dataset for this measurement consists of the magnitude of $S_{21}$ for both qubits, for multiple VNA powers.
We then implement a global fit over this entire dataset, with fit parameters $\Omega_A, \Omega_B,$ and $\eta$ (other parameters include frequency and magnitude offsets).
We constrain the drive strength $\Omega$ by requiring it to be a device-dependent proportionality constant times the square root of the VNA power.
The fit result consists of the solid black curves in Fig.~\ref{fig:characterization}c, and yields a transmission efficiency $\eta^2 = 0.96 \pm 0.01$.

\begin{figure*}[t]
	\centering
	\includegraphics{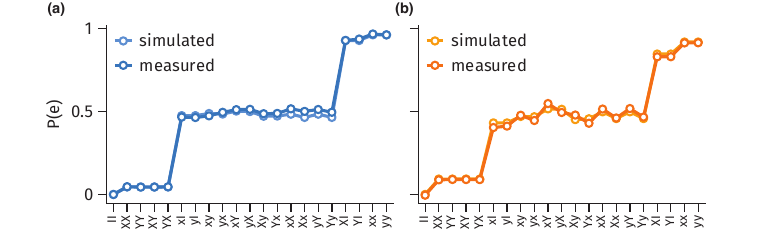}
	\caption{
        \textbf{Comparing simulated and measured \textit{AllXY} measurements.}
        (\textbf{a, b}) Measured and simulated \textit{AllXY} sequence results for qubits A and B respectively.
        Note that for each qubit, the excited state probability is assigned by projecting the readout signal onto a linear scale.
        This scale runs between 0 (corresponding to the ground state) and the excited state probability that can be achieved given the applied $\pi$-pulse and the relaxation time of the qubit.
		}
	\label{suppfig:allXY}
\end{figure*}

\section{Qubit Tuneup}

\subsection{Pulse Calibration}
\label{app:Pulse Tuneup}
The qubits are calibrated using standard techniques. 
The resonant frequency of the qubit is found using pulsed spectroscopy. 
An amplitude Rabi measurement is used to tune up a $\pi$-pulse, followed by a DRAG multiplier calibration using the method described in \cite{reed2013entanglement}. 
Pulse trains are used to fine-tune the amplitudes for $\pi$ and $\pi/2$ rotations. 
All rotations are implemented using 16 ns long pulses with a single-period cosine-square envelope.

\paragraph{Diagnostics} An \textit{AllXY} measurement is used as a diagnostic to check for miscalibrated rotations \cite{reed2013entanglement}. 
With the pulse duration a non-negligible fraction of the qubit lifetime, the rotations are expected to be imperfect. 
The shape of the \textit{AllXY} plot is compared to a master equation simulation of the measurement to confirm if the rotations are limited only by the relaxation and dephasing time of the qubit (Fig.~\ref{suppfig:allXY}). 

\subsection{Readout Optimization}
\label{app:Qubit Readout}

\begin{figure}[t]
    \centering
    \includegraphics{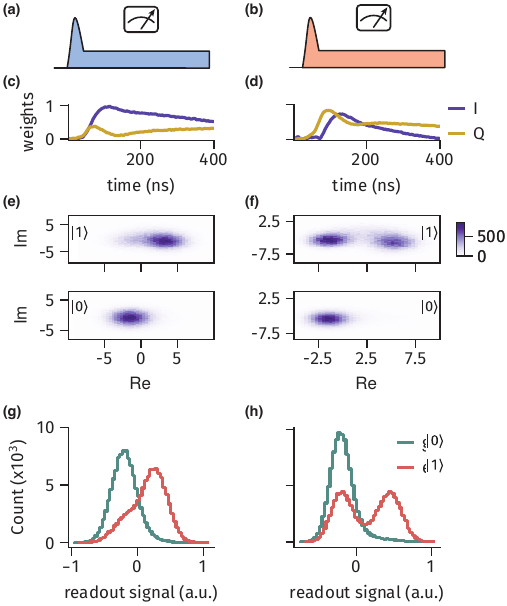}
    \caption{
    \textbf{Qubit readout.}
    (\textbf{a, b}) Pulse diagram for a slingshot readout pulse: an initial high-amplitude segment is added to a square pulse to speed up the resonator ring-up process.
    (\textbf{c, d}) Integration weights for the I and Q quadratures during the readout process for qubits A and B respectively.
    (\textbf{e, f}) Histograms of the integrated readout signal corresponding to the qubit prepared in the ground and excited states. The I and Q quadratures are referred to as the real and imaginary signals.
    (\textbf{g, h}) Histograms shown in (e, f) rotated along the optimal discrimination axis and plotted as one-dimensional histograms.
    }
    \label{suppfig:Readout}
\end{figure}

The qubits are measured using standard dispersive readout. 
The lifetimes of the qubits are on the order of the fastest reported dispersive qubit measurements \cite{walterRapidHighFidelitySingleShot2017a, sunadaFastReadoutReset2022b}. 
This results in a poor signal-to-noise ratio (SNR) between ground and excited state signals. 
We use standard techniques to improve the SNR, enabling sufficient averaging within a reasonable time frame so that readout error is negligible. 
Fig.~\ref{suppfig:Readout} illustrates these techniques and the resulting readout SNR.

\paragraph{Slingshot readout} To speed up the ring-up process of the readout resonator, a square readout pulse is modified to include an initial high amplitude segment \cite{walterRapidHighFidelitySingleShot2017a}.

\paragraph{Non-linear integration weights} SNR is further improved by using non-linear integration weights to demodulate the readout signal \cite{gambettaProtocolsOptimalReadout2007}. 
The optimal integration weights are calculated by subtracting the resonator trajectories corresponding to the qubit being measured in the ground and excited states respectively.
These weights enhance SNR by emphasizing the ring-down, where state contrast is maximal, and suppressing the ring-up, where it is minimal.

\subsection{Drive Calibration}
\label{app: Drive Calibration}
\paragraph{Amplitude} To calibrate the strength of the drive, we measure resonant Rabi oscillations as a function of the drive amplitude. 
The measured oscillation frequency as a function of the amplitude is fit to a linear model to extract a scaling between amplitude and Rabi frequency. 

\paragraph{Phase} For entanglement stabilization, the propagation phase acquired by photons traveling from the upstream qubit to the downstream qubit must be calibrated out.
The steady-state concurrence is measured as a function of the phase of the downstream qubit drive, while the phase of the upstream qubit drive is fixed. 
The phase of the downstream qubit drive that maximizes the concurrence is selected.

\section{Analysis}
\label{app:analysis}
\subsection{State reconstruction}
\label{app:state reconstruction}

Quantum state tomography is used to reconstruct the state of a system by measuring the expectation values of a complete set of measurement operators~\cite{paris2004quantum}.
The measurement operator is determined by a pre-rotation that sets the measurement axis. 
This is accomplished by rotating the state so that the component along a chosen axis is mapped onto the z-axis, which is then measured using standard dispersive readout.
Although dispersive readout alone results in a measurement of the expectation value of $\sigma_z$, preceding it with appropriate rotations allows measurement of the expectation values of $\sigma_x$ and $\sigma_y$, providing sufficient information for reconstructing the state.
Because of the short qubit relaxation times in this system, rotations do not perfectly map the $x$ and $y$ components of the state onto the $z$-axis.
Therefore, these measurements do not correspond to the Pauli operators.

To perform state tomography, we first determine the measurement operators by calculating their corresponding Positive Operator-Valued Measure (POVM) elements.
This procedure is called quantum detector tomography~\cite{lundeen_tomography_2009,blumoff_implementing_2016}.
We achieve this as follows: First, we prepare the qubit in a set of quantum states that we assume we can accurately simulate, allowing us to know their density matrices. 
Next, we measure these states to obtain the expectation values of the measurement operators that we wish to determine.
Finally, we use the expectation values and density matrices to calculate the POVMs corresponding to the measurement operators.

Knowing the POVMs, we can finally measure unknown quantum states and reconstruct their density matrices.
Since this cycle of detector and state tomography relies on the assumption that the set of simulated states matches the experimentally prepared set of states with high fidelity, we perform a series of self-consistency checks. 
We reliably find a high degree of agreement between the reconstructed and simulated states, indicating a small systematic reconstruction error that is then used to assign error bars to the concurrence.
This process is detailed in the following sections.

\subsection{Detector Tomography}
\label{app:detector tomography}

\begin{figure}[t]

    \centering
    \includegraphics{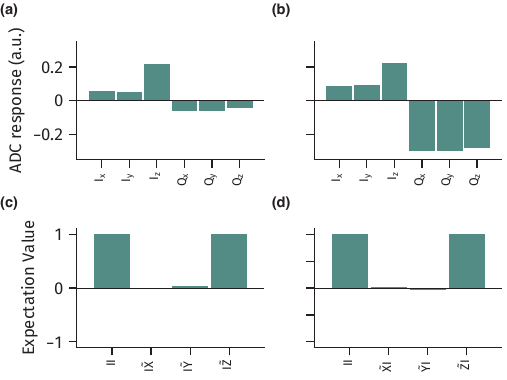}
    \caption{
    \textbf{Measuring cardinal states for detector tomography.}
    (\textbf{a}, \textbf{b}) Average values of the coordinates $(I_i, Q_i)$ (where $i \in \{\tilde{X}, \tilde{Y}, \tilde{Z}\}$ defines the measurement operator) for qubits A and B respectively, obtained by measuring the state $\ket{-\tilde{\mathrm{z}}}_{A,B}$ described in Appendix~\ref{app:detector tomography}.
    (\textbf{c}, \textbf{d}) Expectation values corresponding to the coordinates shown in (a, b).
    }
    \label{suppfig:POVM Reconstruction}
\end{figure}

Prior to performing the experiment, we use detector tomography to characterize the single-qubit measurement operators. 
Detector tomography requires knowing the density matrices of a set of measured states in order to calculate the POVM characterizing the measurement operator.
Therefore, we simulate a set of two-qubit states $\rho^s$, where the state index $s$ denotes the following states: the ground state ($\ket{-\tilde{\mathrm{z}}}_{A,B}$); the state obtained after applying a $\pi$-pulse ($\ket{+\tilde{\mathrm{z}}}_{A,B}$); states obtained after $\pi/2$ and $-\pi/2$-pulses along the x and y axes ($\ket{\pm\tilde{\mathrm{x}}}_{A,B}, \ket{\pm\tilde{\mathrm{y}}}_{A,B}$).
We refer to these as the cardinal states; we use symbols with a tilde to indicate that the states were prepared imperfectly.
For the simulated states to faithfully represent the experimentally prepared states, the following assumptions must be true: 
(i) the qubit is in the ground state before a pulse is applied to prepare a cardinal state, and (ii) relaxation ($T_1$) is the dominant loss mechanism, allowing us to predict state evolution faithfully using master equation simulations.
With these assumptions, we numerically evolve the qubit state using pulses that are identical to those used in the experiment to prepare the cardinal states.
This provides us with the density matrices for the cardinal states.

Next, we experimentally prepare the qubits in one of the cardinal states $\rho^s$ and obtain the expectation values of the single-qubit measurement operators.
Although the cardinal states are defined as two-qubit states, with both qubits prepared in the same single-qubit state (for example, $\ket{-\tilde{\mathrm{z}}}_{A,B} = \ket{-\tilde{\mathrm{z}}}_{A} \otimes \ket{-\tilde{\mathrm{z}}}_{B}$), we use detector tomography to characterize single-qubit measurement operators only. 
Because the two qubits can be read out independently, the two-qubit measurement operators can then be constructed as tensor products of single-qubit measurement operators (described in the next section).
Each single-qubit measurement operator is described by a pre-rotation that sets the measurement axis, followed by a dispersive readout.
We label the measurement operators by $\tilde{X}$, $\tilde{Y}$, and $\tilde{Z}$: $\tilde{X}$ consists of a pre-rotation along the y axis followed by readout, $\tilde{Y}$ consists of a pre-rotation along the x axis followed by readout, and $\tilde{Z}$ consists of an idle pulse of the same duration as the pre-rotations, followed by readout.
Similar to the cardinal state labels, we use symbols with tilde to indicate that the measurement operators do not correspond to the Pauli operators.
Dispersive measurement of a single qubit using standard heterodyne detection yields a coordinate in complex space $(I_i, Q_i)$, where $i \in \{\tilde{X}, \tilde{Y}, \tilde{Z}\}$ defines the measurement operator by setting the measurement axis as discussed above.
This coordinate is used to calculate the expectation value $d_i^s$ for the single-qubit operator defined by $i$ using
\begin{equation}
    d_{i}^s \;=\; 1 - 2 \langle p_{i}^s \rangle  
    \label{eq:onequbitexpval}
\end{equation}
where 
\begin{equation}
    p_{i}^s = \frac{ (I_{i}^s - I_{i}^-)(I_{i}^+ - I_{i}^-) 
         + (Q_{i}^s - Q_{i}^-)(Q_{i}^+ - Q_{i}^-)}
         {(I_{i}^+ - I_{i}^-)^2 + (Q_{i}^+ - Q_{i}^-)^2}.
    \label{eq:probfromIQ}
\end{equation}
$(I_i^+, Q_i^+)$ and $(I_i^-, Q_i^-)$ denote the coordinates of the two cardinal states corresponding to the measurement axis $i$.
Eq.~\ref{eq:probfromIQ} can be understood as follows. 
Each qubit measurement yields a coordinate $(I_i, Q_i)$ that is projected onto a line connecting the two cardinal states coordinates, $(I_i^+, Q_i^+)$ and $(I_i^-, Q_i^-)$, along the measurement axis.
This projection yields a probability $p_i$ that is converted into an expectation value $d_{i}$ using Eq.~\ref{eq:onequbitexpval}.
As an illustration, Fig.~\ref{suppfig:POVM Reconstruction} shows the measured coordinates and the calculated expectation values for the cardinal state $\ket{-\tilde{\mathrm{z}}}_{A,B}$.

The next step is to relate the measured expectation value of a measurement operator to its POVM elements.
A binary measurement can be described by two POVM elements $\{E_i , \mathbb{I} - E_i\}$ that sum up to identity, where $i \in \{\tilde{X}, \tilde{Y}, \tilde{Z}\}$ defines the measurement operator by setting the measurement axis.
For a given state $\rho$, the probabilities of the two outcomes are $\mathrm{Tr}[\rho E_i]$ and $\mathrm{Tr}[\rho (\mathbb{I}-E_i)]$ respectively.
The expectation value can therefore be written as the weighted average
\begin{equation}
    d_i = \alpha_i \,\mathrm{Tr}[\rho E_i] + \beta_i \,\mathrm{Tr}[\rho (\mathbb{I}-E_i)],
    \label{eq:single qubit exp values}
\end{equation}
where $\alpha_i$ and $\beta_i$ are the outcome values assigned to the two POVM elements.

Eq.~\ref{eq:single qubit exp values} relates the measured expectation value $d_{i}^s$ obtained from Eq.~\ref{eq:onequbitexpval}, the simulated density matrix $\rho^s$, and the POVM elements $\{E_i , \mathbb{I} - E_i\}$.
We perform a convex optimization on Eq.~\ref{eq:single qubit exp values} using the Python package \texttt{cvxpy}~\cite{diamond2016cvxpy} to obtain the POVM elements describing the single-qubit measurement operators.

\subsection{State tomography}
Detector tomography provides the POVMs characterizing the measurement operators for each qubit.
Next, we perform the entanglement stabilization experiment by applying drives to both qubits until they reach the steady-state.
We then turn off the drives and immediately measure the expectation values of two-qubit operators.
The two-qubit operators are indexed using $i$ and $j$, where $i, j \in \{\tilde{X}, \tilde{Y}, \tilde{Z}\}$ denote the measurement operators for qubits A and B respectively.
To measure the expectation value of a two-qubit operator, we simultaneously measure both qubits, acquiring all pairwise products of $I_i$, $Q_i$, $I_j$, and $Q_j$.
These quantities are used to calculate the expectation value $d_{i,j}$ for a two-qubit operator defined by $i$ and $j$ using
\begin{align}
    d_{i, j} \;&=\; 
    (1 - 2 \langle p_{i} \rangle - 2 \langle p_{j} \rangle + 4\langle  p_i p_j \rangle ).
\label{eq:twoqubitexpval}
\end{align}
The product of probabilities in Eq.~\ref{eq:twoqubitexpval} can be written as

\begin{widetext}
\begin{equation}
\begin{aligned}
\label{eq:probproduct}
\langle p_{i}^s p_{j}^s \rangle &= \frac{1}{D_{i} D_{j}} \Bigg( 
        \Big[ \langle I_{i}^s I_{j}^s \rangle - \langle I_{i}^s \rangle I^{-}_{j} - \langle I_{j}^s \rangle I^{-}_{i} + I^{-}_{i} I^{-}_{j} \Big] \, \Delta I_{i}^s \Delta I_{j}^s \\
        &\quad + \Big[ \langle I_{i}^s Q_{j}^s \rangle - \langle I_{i}^s \rangle Q^{-}_{j} - \langle Q_{j}^s \rangle I^{-}_{i} + I^{-}_{i} Q^{-}_{j} \Big] \, \Delta I_{i}^s \Delta Q_{j}^s \\
        &\quad + \Big[ \langle Q_{i}^s I_{j}^s \rangle - \langle Q_{i}^s \rangle I^{-}_{j} - \langle I_{j}^s \rangle Q^{-}_{i} + Q^{-}_{i} I^{-}_{j} \Big] \, \Delta Q_{i}^s \Delta I_{j}^s \\
        &\quad + \Big[ \langle Q_{i}^s Q_{j}^s \rangle - \langle Q_{i}^s \rangle Q^{-}_{j} - \langle Q_{j}^s \rangle Q^{-}_{i} + Q^{-}_{i} Q^{-}_{j} \Big] \, \Delta Q_{i}^s \Delta Q_{j}^s
        \Bigg)
\end{aligned}
\end{equation}
\end{widetext}

\begin{figure*}[t]
    \centering
    \includegraphics{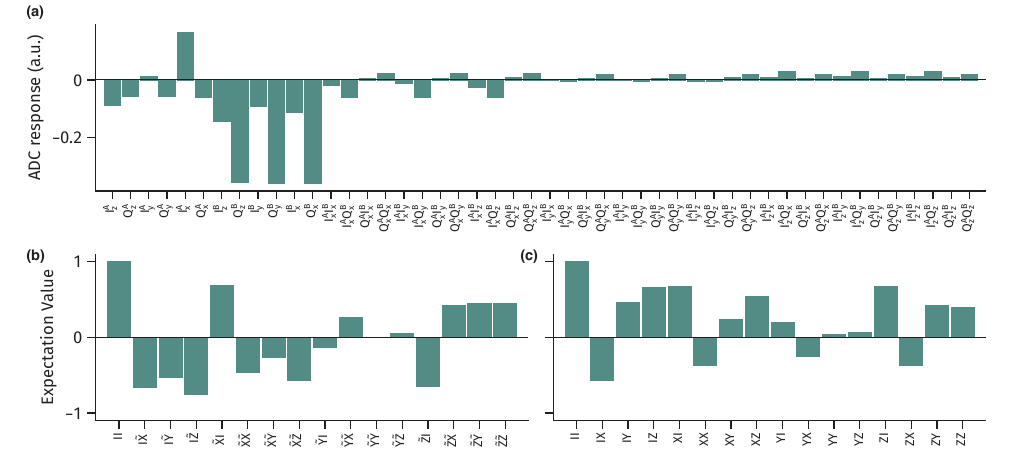}
    \caption{
    \textbf{State Tomography.}
     (\textbf{a}) Average values of the measured pairwise products of $I_i$, $Q_i$, $I_j$, and $Q_j$ (where $i, j \in \{\tilde{X}, \tilde{Y}, \tilde{Z}\}$ denote the measurement operators for qubits A and B respectively) for an unknown two-qubit state.
     (\textbf{b}) Expectation values of the two-qubit measurement operators calculated from the pairwise products shown in (a).
     (\textbf{c}) Pauli bars represent the density matrix reconstructed from the expectation values shown in (b).
    }
    \label{suppfig:state recon}
\end{figure*}

where $D_{i} = ( I_{i}^{+} - I_{i}^{-})^{2} + ( Q_{i}^{+} - Q_{i}^{-} )^{2}$, $\Delta I_{i} = I_{i}^{+} - I_{i}^{-}$, and $\Delta Q_{i} = Q_{i}^{+} - Q_{i}^{-}$.
Next, we relate the expectation value $d_{i, j}$ to the POVM elements of the two-qubit measurement operator.
Since the two-qubit POVMs are simply tensor products of the single-qubit POVMs, Eq.~\ref{eq:single qubit exp values} can be extended straightforwardly to the two-qubit case:
\begin{equation}
    \begin{aligned}
        d_{i, j} &= \alpha_{i}\alpha_{j} \, \mathrm{Tr}\!\left[ \rho \,(E_{i} \otimes E_{j}) \right] 
        + \alpha_{i}\beta_{j} \, \mathrm{Tr}\!\left[ \rho \,(E_{i} \otimes (\mathbb{I} - E_{j})) \right] \\
        &\quad + \beta_{i}\alpha_{j} \, \mathrm{Tr}\!\left[ \rho \,((\mathbb{I} - E_{i}) \otimes E_{j}) \right]  \\
        &\quad + \beta_{i}\beta_{j} \, \mathrm{Tr}\!\left[ \rho \,((\mathbb{I} - E_{i}) \otimes (\mathbb{I} - E_{j})) \right].
    \end{aligned}
    \label{eq:two qubit exp values}
\end{equation}
With the expectation values $d_{i, j}$ measured, and the single-qubit POVMs calculated using detector tomography, we perform a convex optimization on Eq.~\ref{eq:two qubit exp values}, subject to physicality constraints, to obtain the steady-state $\rho$.
As an illustration, Fig.~\ref{suppfig:state recon} shows the expectation values of the two-qubit measurement operators for an unknown two-qubit state, and the Pauli bars represent the reconstructed state.

It is important to note that the two qubits have independent readout and therefore their measurement operators are described by independent POVMs.
Despite this, the reconstruction procedure captures the off-diagonal terms of the two-qubit density matrix as the correlated information is encoded in the pairwise products of $I_i$, $Q_i$, $I_j$, and $Q_j$ in Eq.~\ref{eq:probproduct}, which are multiplied together before being averaged.

\subsection{Reconstruction Error}
\label{app:Reconstruction Error}
\begin{figure*}[t]
    \centering
    \includegraphics{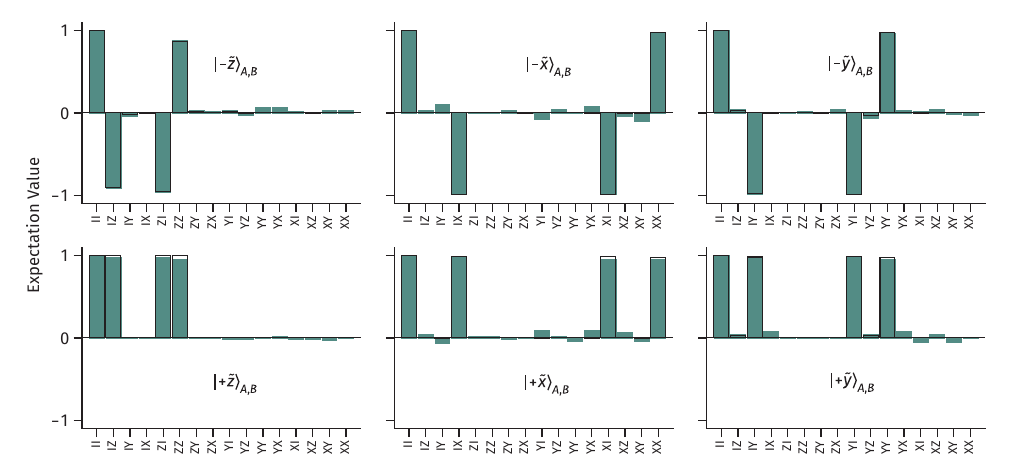}
    \caption{
    \textbf{Reconstructed cardinal states.}
    Expectation values of Pauli operators for the reconstructed (teal) and simulated (black outlined) cardinal states.
    }
    \label{suppfig:Reconstruction Error}
\end{figure*}

The state reconstruction procedure relies on an accurate simulation of the cardinal states that are used to calculate the POVMs.
As a consistency check, we use state tomography to reconstruct the cardinal states, and compare them to the simulated cardinal states. 
We find on average $97\%$ fidelity between the reconstructed and simulated cardinal states (illustrated in Fig.~\ref{suppfig:Reconstruction Error}), indicating a self-consistent cycle of detector and state tomography.

The difference between the simulated and reconstructed states is used to assign a systematic error to reconstructed density matrix elements.
To do this, we perform the following steps.
First, we subtract the single-qubit Pauli operator expectation values for the simulated and reconstructed cardinal states (Fig.~\ref{suppfig:Reconstruction Error}).
This provides us with a set of systematic reconstruction errors for each single-qubit Pauli operator.
Because the six cardinal states form a reasonably unbiased set of single-qubit states, we use this set directly as the error distribution, which is assumed to be Gaussian.
For each two-qubit Pauli-operator product, the corresponding single-qubit error distributions are combined to yield the product error distribution.

For an unknown state, we first reconstruct its density matrix to obtain the two-qubit Pauli operator expectation values. 
We then perform bootstrapping using the error distributions: in each iteration, an error is sampled from the error distribution for every two-qubit Pauli operator product, added to the measured value, and a density matrix is reconstructed from the perturbed set of expectation values.
Repeating this procedure 100 times yields an ensemble of density matrices, from which we obtain a distribution of the concurrence. 
The uncertainty in concurrence is reported as the 68\% confidence interval, corresponding to the 15th and 84th elements of the sorted concurrence values.

This systematic error dominates over the statistical error originating from dispersive readout of the qubits (Fig.~\ref{suppfig:Readout}g,h).
Based on the signal-to-noise ratio of dispersive readout of qubits (Fig.~\ref{suppfig:Readout}g,h), we average each measurement such that the statistical error is negligible compared to the systematic reconstruction error, and can thus be ignored.

\subsection{Measurement workflow}
\label{app:averaging}
The qubits are tuned up based on the routine outlined in Appendix~\ref{app:Pulse Tuneup} once a day.
Given the low readout signal-to-noise ratio, the tune-up routine takes approximately half an hour.
It is therefore impractical to interleave measurements with the tune-up routine.

Low readout readout signal-to-noise ratio requires long averaging times for each measurement. 
As a result, each sweep over $\Omega_A$ (Fig.~\ref{fig:CQA}a and Fig.~\ref{fig:landscape}b) takes approximately 1.5 hours.
This sets a practical limit for the amount of measurements that can be performed. 
Such long acquisition times may allow qubit parameters to drift, reducing the reconstructed state fidelity.

\subsection{Simulating states for detector tomography}
\label{app: Simulations}
Quantum detector tomography requires accurate simulation of the cardinal states defined in Appendix~\ref{app:detector tomography}.
We use the Lindblad Master Equation to evolve the state of the transmon under the effect of the rotation pulses used to prepare the cardinal states in experiment. 
We model the transmon as a three-level system to account for the anharmonicity.
The system parameters used in the simulation are listed in table~\ref{tab:transmon parameters}.
The self-consistency of this cycle of detector and state tomography is confirmed by reconstructing the cardinal states, as described in Appendix~\ref{app:detector tomography}.
As a preliminary check, however, we perform an \textit{AllXY} measurement \cite{reed2013entanglement} on each qubit. We find excellent agreement with simulations (Fig.~\ref{suppfig:allXY}), confirming that the pulses are accurately modeled. 
The small discrepancy between the measured and simulated \textit{AllXY} sequence is consistent with the reconstruction error discussed in Appendix~\ref{app:detector tomography}.
Thus, this comparison serves as a consistency check that the uncertainty calculated from the reconstruction error faithfully captures the error in the tomography process.

\subsection{Modeling steady-state concurrence}
\label{app:modeling}
To model the experimentally measured steady-state concurrence, we numerically simulate the steady-state using the Lindblad Master Equation described by Eq.~\ref{eq-app:Hamiltonian}. 
To account for the anharmonicity of the transmon, we replace the spin-1/2 operators with bosonic ladder operators and include the first three levels in the model.

\paragraph{Drive-induced frequency shift} The drive strength that results in the maximum concurrence is determined by the detuning between the qubit frequencies.
We first simulate the steady-state concurrence using the detuning between the measured qubit frequencies listed in Table~\ref{tab:transmon parameters} and find that the peak of the simulated concurrence occurs a few megahertz away from the measured peak (Fig.~\ref{suppfig:starkshift}a).
This difference is illustrated by the solid and dashed curves in Fig.~\ref{suppfig:starkshift}a.
This is indicative of a shift in the transmon frequencies that is not predicted by master equation simulations described in Appendix~\ref{app: Simulations}. 
One possibility is imperfect modeling of the transmon Hamiltonian, for example, by neglecting higher harmonics in the Josephson potential~\cite{willsch2024observation} which are known to cause corrections to bare qubit frequencies.
Another plausible contribution is an ac Stark shift due to the off resonant drives~\cite{schuster2005ac}.
Predicting this shift from independent theory and experiments is not straightforward in our setup.
For example, non-uniform transmission in the microwave lines results in frequency-dependent power reaching the device.
We routinely observe small deviations from theoretical predictions in a range of different experiments, such as qubit tuneup.

Since the shifted qubit frequencies are encoded in the drive strength that yields maximum concurrence, we use it here as an empiric measure of the shift. 
We add the following term to the Hamiltonian
\begin{equation}
    \hat{H}_\mathrm{shift} = (\lambda\, \Omega^2 / \Delta) \, \hat{a}^\dagger\hat{a},
\label{eq:starkshift}
\end{equation}
treating it analogously to an ac Stark shift, where $\Omega$ is the drive strength, $\Delta$ is the drive detuning from the resonant frequency, and $\lambda$ is a device-dependent parameter.
With $\lambda_A$ and $\lambda_B$ as the only free parameters, we fit the concurrence as a function of drive strength and achieve high fidelity between the measured and simulated steady-states.
These fits are represented by the solid curves in Fig.~\ref{fig:landscape}b.
\paragraph{Concurrence landscape}
As described above, the drive-induced qubit frequency shift is dependent on the drive frequency. 
Each fit in Fig.~\ref{fig:landscape}b yields a pair of drive frequency shift parameters ($\lambda_A, \lambda_B$) for the corresponding drive frequency.
This pair of parameters is used to plot the corresponding concurrence contour plot in Fig.~\ref{fig:landscape}a.
Therefore, these contour plots show how the concurrence landscape would be if these parameters remained constant for all detunings $\varepsilon$.
Fig.~\ref{suppfig:starkshift}(b, d) show the difference between the concurrence landscape had there been no drive-induced frequency shift.

\subsection{Sensitivity to drive detuning} 
\label{app:sensitivity}
While the measurements for Fig.~\ref{fig:CQA}a and Fig.~\ref{fig:CQA}b were performed with identical drive parameters, there is a slight discrepancy in the steady-state concurrence.
The peak concurrence in Fig.~\ref{fig:CQA}a is 0.135, while the average concurrence in the plateau of Fig.~\ref{fig:CQA}b is 0.173.
These measurements were performed at different times, and were therefore susceptible to drifts in experimental parameters, including qubit frequency and coherence, as well as the power stability of the microwave electronics that is used for measurement; we typically observe drifts in these parameters over the course of hours and days, impacted by environmental conditions in the laboratory.
These observations are consistent with the reported fluctuations.
As mentioned in \cref{app:averaging}, the long integration times of the experiments make it challenging to recalibrate the experiment frequently.

\begin{figure*}[t]
	\centering
	\includegraphics{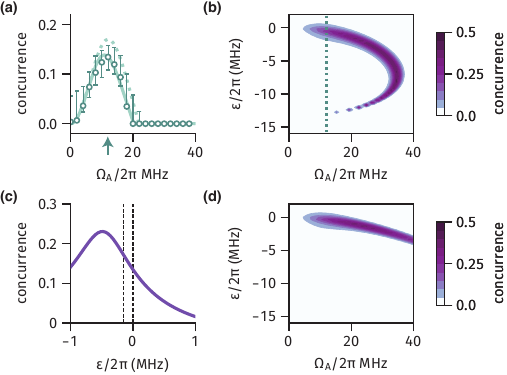}
	\caption{
        \textbf{Drive-induced frequency shift and sensitivity to detuning.}
        (\textbf{a}) Measured steady-state concurrence as a function of drive strength $\Omega_A$ for $\Omega_B/\Omega_A = 1$ and $\varepsilon/2\pi = 0~\mathrm{MHz}$ (circular markers). This is the same data as Fig.~\ref{fig:CQA}a. The solid line represents the master equation fit, taking into account a drive-induced frequency shift described by Equation \ref{eq:starkshift}.
        The dashed teal line shows master equation simulation without taking into account the frequency shift.
        (\textbf{b}) Simulated stead-state concurrence as a function of the drive strength $\Omega_A$ and the drive detuning $\varepsilon$ for $\Omega_B/\Omega_A = 1$. The simulation takes into account the drive-induced frequency shift obtained from the fit in (a). The dashed teal is at the drive strength that yields the maximum concurrence in (a), as pointed out by the teal arrow.
        (\textbf{c}) A line cut from (c) illustrated by the dashed teal arrow. 
        This represents the drive strength that yields the maximum concurrence in (a). The black dashed line at zero detuning represents the maximum concurrence in (a) ($\conc=0.135$). The offset dashed line is $160~\mathrm{kHz}$ away, and corresponds to $\conc=0.173$.
        (\textbf{d}) The same simulation as (b) without taking into the drive-induced frequency shift. 
		}
	\label{suppfig:starkshift}
\end{figure*}

Fig.~\ref{suppfig:starkshift}c shows how this discrepancy can be explained by a $160~\mathrm{kHz}$ relative qubit frequency drift between the two measurement times.
Fig.~\ref{suppfig:starkshift}c also shows that the concurrence value shown in Fig.~\ref{fig:CQA}a is less than the peak concurrence attainable at this drive strength.
This is due to the following.
Drive-induced frequency shifts are not straightforward to predict in this setup (Appendix~\ref{app:modeling}).
Therefore, it is only after fitting concurrence as a function of drive strength (solid curves in Fig.~\ref{fig:landscape}b) that we find the actual detuning due to the drive-induced frequency shifts.
In principal, one could optimize over both drive strength and detuning to map out the concurrence landscape. 
This optimization was not performed in this experiment as a better protocol was discovered that resulted in a steady-state concurrence higher than what could be achieved using the CQA protocol.

\section{Deriving the Master Equation}
\label{app:slh}
The derivation of the Lindblad Master Equation for a cascaded network of two qubits is described in \cite{gardiner_driving_1993a, carmichael_quantum_1993a}. The Markov approximation is used to trace out the waveguide modes in order to obtain a master equation for the two-qubit system. Alternatively, we can employ the SLH formalism to derive the master equation for this system \cite{combesSLHFrameworkModeling2017}. 
Each component in this network is specified by a triple $(S, L, H)$ that describes how it interacts with input and output fields. 
$S$ is the scattering matrix for the component, $L$ contains its coupling to external fields, and $H$ is the Hamiltonian. 
The cascaded network of two qubits coupled to a unidirectional waveguide can be modeled using three components in series configuration: qubit A, followed by a fictitious beam splitter, followed by qubit B. 
The beam splitter has a probability $\eta^2$ of allowing a photon to propagate through from qubit A to qubit B, allowing us to model photon loss in the waveguide.
The SLH triple for the $i$-th qubit is given by 
\begin{align*}
    S &= \mathbb{I} \\
    L &= \sqrt{\gamma} \begin{pmatrix}
        \hat{\sigma}_{i}^- \\
        0
        \end{pmatrix} \\
    \hat{H} &= \frac{\Delta}{2} \hat{\sigma}_{i}^z + \frac{\Omega}{2} \hat{\sigma}_{i}^x.
\end{align*} 
The Hamiltonian is in the frame of the drive which is detuned from the qubit frequency by $\Delta$. The SLH triple for the beam splitter is given by
\begin{align*}
    S &= \begin{pmatrix}
        \eta & -\sqrt{1-\eta^2} \\
        \sqrt{1-\eta^2} & \eta
        \end{pmatrix} \\
    L &= 0 \\
    \hat{H} &= 0
\end{align*} where $\eta^2$ is the probability that a photon propagates through from qubit A to qubit B. 
For two components connected in series, the product rule used to calculate the SLH triple for the network is
\begin{align}
    &(S_2, L_2, H_2) \triangleleft (S_1, L1, H_1) = \label{eq:SLH_series}\\
    &\Big(S_2 S_1, L_2 + S_2 L1, H_1 + H_2 + \frac{1}{2i}(L_2^{\dagger} S_2 L_1 - L_1^{\dagger} S_2^{\dagger} L_2)\Big).\nonumber
\end{align}
Using Eq.~\ref{eq:SLH_series}, we obtain the following SLH triple for our cascaded system:

\begin{equation}
    \Bigg( 
            \begin{pmatrix}
            \sqrt{\eta} & \sqrt{1-\eta} \\
            -\sqrt{1-\eta} & \sqrt{\eta}
            \end{pmatrix},
            \sqrt{\gamma} \begin{pmatrix}
                \sqrt{1-\eta^2}\hat{\sigma}_{A}^{-} \\
                \eta \hat{\sigma}_{A}^{-} + \hat{\sigma}_{B}^{-}
                \end{pmatrix},
            \hat{H}
    \Bigg) 
\end{equation}
where the Hamiltonian is given by
\begin{align}
    \hat{H} &= \Big(\frac{\Delta_A}{2} \hat{\sigma}_{A}^{z} + \frac{\Omega_A}{2} \hat{\sigma}_{A}^{x} \Big) + \Big (\frac{\Delta_B}{2} \hat{\sigma}_{B}^{z} + \frac{\Omega_B}{2} \hat{\sigma}_{B}^{x} \Big) \\
    &- i \frac{\eta~\gamma}{2} \Big( \hat{\sigma}_{A}^{-}\hat{\sigma}_{B}^{+} - \hat{\sigma}_{A}^{+}\hat{\sigma}_{B}^{-} \Big)\nonumber
\end{align}
and the collapse operators are 
\begin{align}
    \hat{c}_1 &= \eta \hat{\sigma}^{-}_A + \hat{\sigma}^{-}_B \nonumber \\
    \hat{c}_2 &= \sqrt{1-\eta^2} \hat{\sigma}^{-}_A. 
\end{align}
Finally, taking $\Omega_B=\Omega_A$ and $\Delta_B=-\Delta_A$, we obtain 
\begin{equation}
    \hat{H} = \frac{\Omega}{2}(\hat{\sigma}^{x}_A + \hat{\sigma}^{x}_B) + \frac{\Delta}{2}(\hat{\sigma}^{z}_A -\hat{\sigma}^{z}_B) + i\eta\frac{\gamma}{2}(\hat{\sigma}^{+}_A \hat{\sigma}^{-}_B - \mathrm{h.c.})
    \label{eq:2qHamiltonian} 
\end{equation}
and
\begin{align}
    \hat{c}_1 &= \eta \hat{\sigma}^{-}_A + \hat{\sigma}^{-}_B \nonumber \\
    \hat{c}_2 &= \sqrt{1-\eta^2} \hat{\sigma}^{-}_A. \label{eq:c2}
\end{align}

\section{Stabilizing entanglement using the Coherent Quantum Absorber scheme}
\label{app:CQA intro}
The Coherent Quantum Absorber (CQA) scheme is a driven-dissipative entanglement stabilization protocol in which an auxiliary quantum system acts as a perfect absorber of the primary system such that the composite system relaxes into an entangled state.
We take the composite system to be two driven qubits coupled to a unidirectional waveguide.
The drives are applied at the center frequency of the qubits.
The Hamiltonian for this system is derived in Appendix~\ref{app:slh}.
For the following derivation, we assume that there is no waveguide loss, and hence only a single loss operator $\hat{c}=\hat{\sigma}^{-}_A + \hat{\sigma}^{-}_B$.
To find a pure state of this system, we look for a dark state (a state that gives zero when the collapse operator is applied to it). 
This describes a state of the system in which no photons propagate beyond the two qubits.
One can see that there is a dark state of this system that is also an eigenstate of the Hamiltonian with a zero eigenvalue, and therefore a steady state of the system, given by
\begin{equation}
    \ket{\psi_0} = \ket{00} + \frac{\sqrt{2}\Omega}{2\Delta - i\gamma}\ket{S},\label{eq:psi0-delta}
\end{equation}
up to a normalization constant, where $\ket{S}=(|01\rangle-|10\rangle)/\sqrt{2}$ is the singlet state \cite{stannigel_drivendissipative_2012a, motzoi_backactiondriven_2016}.

\section{Deriving symmetry conditions for optimal entanglement} 
\label{app:TMS}

\paragraph{Derivation} For a given experimental setup, the qubit frequencies $\omega_i$ and waveguide couplings $\gamma_A$, $\gamma_B$ are fixed, while the drive amplitudes $\Omega_A$, $\Omega_B$ and the drive detuning $\epsilon$ (from the center frequency) may be varied. 
In the rotating frame of the drive, the system is described by the following master equation:
\begin{equation}
    \begin{aligned}
        \hat{H} &= \hat{H}_A + \hat{H}_B+\hat{H}_{s}\\
        &=\Omega_A \hat{\sigma}_A^x + (\Delta+\epsilon)\hat{\sigma}_A^z + \Omega_B \hat{\sigma}_B^x - (\Delta -\epsilon) \hat{\sigma}_B^z \\
        &\quad - \frac{i\sqrt{\gamma_A \gamma_B}}{2} (\hat{\sigma}_A^- \hat{\sigma}^+_B - h.c.) 
        \\
        \hat{L} &= \sqrt{\gamma_A} \hat{\sigma}_A^- + \sqrt{\gamma_B} \hat{\sigma}_B^- 
        \label{eq-app:Hamiltonian}
    \end{aligned}
\end{equation}

We begin our analysis under the assumption that the qubit detuning $\Delta$ dominates over the dissipation rates, i.e., $\gamma_A, \gamma_B \ll \Delta$. 
This condition is met in our experiment, where $\gamma \sim 1~\text{MHz}$ and $\Delta \sim 10~\text{MHz}$.

To analyze the dynamics in the Rabi frame, we perform a unitary rotation about the $y$-axis on each qubit: $\hat{U}_A = \hat{R}_A^y(\theta_A) = \exp(-i\theta_A \hat{\sigma}_A^y/2)$ and $\hat{U}_B = \hat{R}_B^y(-\theta_B) = \exp(i\theta_B \hat{\sigma}_B^y/2)$. 
This rotates the local Hamiltonian $\hat{H}_A$ and $\hat{H}_B$ into their respective eigenbasis. 
The transformed Hamiltonian becomes:
\begin{equation}
    \hat{H}'= \Tilde{\Delta}_A \hat{\tau}_A^z -\Tilde{\Delta}_B \hat{\tau}_B^z + \hat{H}_{\rm diss}, \label{app-eq:after-rotation}
\end{equation}
where $\tilde{\Delta}_A = \sqrt{\Omega_A^2 + (\Delta + \epsilon)^2}$ and $\tilde{\Delta}_B = \sqrt{\Omega_B^2 + (\Delta - \epsilon)^2}$ are the local energy splittings in the rotated frame. 
The term $\hat{H}_{\mathrm{diss}} \propto \sqrt{\gamma_A \gamma_B}$ represents the cascaded interaction after the rotation, which we treat perturbatively. 
The transformed Pauli operators are denoted by $\hat{\tau}_{A(B)}^\alpha$, and
the rotation angles are given by $\sin \theta_{A(B)} = \Omega_{A(B)} / \tilde{\Delta}_{A(B)}$. 
The transformed jump operator can be divided into three terms $\hat{L}'=\hat{L}_+ + \hat{L}_-+ \hat{L}_z$:
\begin{align}
    \hat{L}_- &= \sqrt{\gamma_A} (\cos \theta_A+1)\hat{\tau}_A^- +\sqrt{\gamma_B} (\cos \theta_B-1) \hat{\tau}_B^+, \\
    \hat{L}_+ &= \sqrt{\gamma_B} (\cos \theta_B+1)\hat{\tau}_B^- +\sqrt{\gamma_A} (\cos \theta_A-1) \hat{\tau}_A^+, \\
    \hat{L}_z &= \sqrt{\gamma_A} \sin \theta_A \hat{\tau}^z_A - \sqrt{\gamma_B} \sin \theta_B \hat{\tau}^z_B.
\end{align}
We group terms by their frequencies in the interaction picture $\hat{L}_+(t) \approx \hat{L}_+ e^{i\tilde{\Delta}t}, \; \hat{L}_-(t) \approx \hat{L}_- e^{-i\tilde{\Delta}t}$, and $ \hat{L}_z(t) \approx \hat{L}_z$ where we eliminate the fast rotating cross terms.
The two jump operators $\hat{L}_-$ and $\hat{L}_+$ mimic two-mode squeezing dissipators---a phenomenon referred to as `synthetic squeezing' \cite{govia_stabilizing_2022}.
To ensure these dissipators stabilize the same entangled state, we match their squeezing strengths by setting
\begin{equation}
    \tanh r \equiv  \frac{\sqrt{\gamma_B} (1- \cos \theta_B)}{\sqrt{\gamma_A} (1+ \cos \theta _A)} = \frac{\sqrt{\gamma_A} (1- \cos \theta_A)}{\sqrt{\gamma_B}(1+ \cos \theta _B)}.
\end{equation}
This equation is further simplified to obtain the optimal driving strength
\begin{equation}
    \gamma_A \Omega_{A,\mathrm{opt}}^2 = \gamma_B \Omega_{B,\mathrm{opt}}^2. 
    \label{app-eq:drive-optimal}
\end{equation}
The dissipators now take the form
\begin{align}
    \hat{L}_- &= \sqrt{\tilde{\gamma}_-} \left( \cosh r \;\hat{\tau}_A^- +\sinh r\; \hat{\tau}_B^+ \right), \\
    \hat{L}_+ &= \sqrt{\tilde{\gamma}_+} \left( \cosh r \;\hat{\tau}_B^- +\sinh r\; \hat{\tau}_A^+ \right), \\
    \hat{L}_z &= \sqrt{\gamma_A \Omega_A^2} (\hat{\tau}^z_A -  \hat{\tau}^z_B),
\end{align}

where $\tilde{\gamma}_- = 2 \gamma_A \frac{1+\cos \theta_A}{1+\cos\theta_B} (\cos \theta_A+ \cos \theta_B)$ and $\tilde{\gamma}_+ = 2 \gamma_B \frac{1+\cos \theta_B}{1+\cos\theta_A} (\cos \theta_A+ \cos \theta_B)$ are renormalized dissipation rates. Notice that all three dissipators support the same two mode squeezed state
\begin{equation}
    |\psi_{\rm TMS}\rangle = \left(\cosh r |00\rangle + \sinh r|11\rangle\right)/\sqrt{\cosh 2r},
\end{equation}
with $\hat{L}_{\pm} |\psi_{\rm TMS}\rangle = 0$ and $\hat{L}_{\rm z} |\psi_{\rm TMS}\rangle = 0$.

To ensure that the Hamiltonian in Eq.~(\ref{app-eq:after-rotation}) is compatible with the dark state $|\psi_{\rm TMS} \rangle$, we find that up to zeroth order in $\gamma_{A,B}/\Delta$ the frequencies $\tilde{\Delta}_A$ and $\tilde{\Delta}_B$ must be matched. This constrains the drive frequency to
\begin{equation}
    \epsilon_{\mathrm{opt}}  = \frac{\Omega_B^2 - \Omega_A^2}{4\Delta}. \label{app-eq:detuning-optimal}
\end{equation}
Moreover, the relative phase between both drives must be calibrated in order to cancel out the phase acquired by the photon as it propagates from the upstream to the downstream qubit. 
A residual phase offset manifests as a phase in the effective jump operators in the frame of the drive. 
This causes each jump operator to stabilize a different target state, thereby reducing the degree of stabilized entanglement.

\paragraph{Requirement for purity} The non-reciprocal interaction term 
\begin{equation}
    \hat{H}_{\rm diss} = i \sqrt{\gamma_A\gamma_B}(\cos \theta_A - \cos \theta_B) (\sigma_A^+ \sigma_B^+ - h.c.) 
\end{equation}
causes the system to escape the dark state manifold due to a phase flip. Even if we satisfy the two optimal conditions \cref{app-eq:drive-optimal,app-eq:detuning-optimal}, this additional term will eventually generate an incoherent mixture between $| \psi_{\rm TMS} \rangle$ and $|\psi_{\rm Flip}\rangle \sim \cosh r |11 \rangle - \sinh r | 00\rangle$, when asymmetry $\theta_A \neq \theta_B$ exists.
Therefore, the only way to have the system stabilize a pure state is by ensuring that $\theta_A = \theta_B$, which imposes the symmetry required by the CQA scheme.

\paragraph{Physical intuition} Eq.~(\ref{app-eq:detuning-optimal}) ensures that the photons emitted by the two qubits have the same frequency so they can coherently interfere. In the limit of a weak drive, the intensity of photon emission can be derived using the input-output theorem to be $\langle n\rangle = \sqrt{\gamma} \langle \hat{\sigma}^+ \hat{\sigma}^-\rangle \sim \sqrt{\gamma}  \Omega^2$. Requiring the intensity of the photons emitted by the two qubits to be equal leads to Eq.~(\ref{app-eq:drive-optimal}).

\paragraph{Robustness against coupling mismatch} The CQA scheme relies on a hidden time-reversal symmetry to stabilize a pure entangled state \cite{robertsHiddenTimeReversalSymmetry2021}. 
This symmetry lacks robustness against coupling mismatch.
The symmetry associated with `synthetic squeezing', however, can be enforced despite arbitrary coupling mismatch simply by tuning the amplitudes and frequency of the drives.
This makes it a more appealing strategy for implementation as variability in qubit-waveguide coupling strengths is often inevitable in experiment.

\paragraph{Effect of local noise} The degree of stabilized entanglement is limited by local loss. 
This includes qubit relaxation and dephasing (see \cref{tab:transmon parameters}), and $4\%$ transmission loss in the waveguide. 
In \cref{app:slh}, we discuss how having transmission loss in the waveguide is equivalent to having asymmetric qubit-waveguide couplings and additional decay on the upstream qubit. While the asymmetric coupling can be compensated by asymmetric drives, qubit decay causes the system to jump to the vacuum state $\hat \sigma^-_A \ket{\psi_{\rm CQA}} = -\hat \sigma^-_B \ket{\psi_{\rm CQA}} \propto \ket{00}$. The drive and dissipative coupling to the waveguide re-pump the system into the target state $\ket{\psi_{\rm CQA}}$. This leads to a competition between two dissipative process. In our experiment, the intrinsic relaxation time is two orders of magnitude longer than engineered relaxation. Using a perturbative argument, the resulting density matrix can be shown to be $\hat \rho \sim (1- p) \hat \rho_{\rm target} + p \ket{00}\bra{00}$, where $p \approx \gamma_{T_1}/\gamma_{\rm diss}$ is the ratio between two processes. For more details on the effect of loss, see \cite{irfan_loss_2024a}.

\section{Driving the upstream qubit}
\label{app: Upstream Drive}
Applying a drive only on the upstream qubit does not lead to a pure steady state, making an analytical solution non-trivial.
However, to gain insight into how entanglement may be stabilized using this drive scheme, we work in the regime where the drive acts as a perturbation to the two-qubit Hamiltonian coupled to the waveguide, described by
\begin{align}
    \hat{{H}} &= \Omega \sigma_A^x - \frac{i \gamma}{2}(\hat{\sigma}_A^-\hat{\sigma}_B^+ - h.c.) \\
    \hat{{L}} &= \sqrt{\gamma} (\sigma_A^- + \sigma_B^-).
\end{align}
This approximation holds when  $\Omega /\gamma \ll 1$.
In this limit we can assume the probability of a quantum jump is very small ($\gamma \langle (\sigma_A^+ + \sigma_B^+)(\sigma_A^- + \sigma_B^-) \rangle \ll 1$), allowing us to write the effective Hamiltonian as 
\begin{align}
    \hat{{H}}_{\rm {eff}} &= \hat{{H}} - \frac{i}{2} \hat{{L}}^{\dagger} \hat{{L}} \\
    &= \Omega \hat{\sigma}_A^x - i\gamma \hat{\sigma}_A^- \hat{\sigma}_B^+ - \frac{i\gamma}{2} \hat{\sigma}_A^+\hat{\sigma}_A^- - \frac{i\gamma}{2} \hat{\sigma}_B^+\hat{\sigma}_B^-
\end{align}
The effective Hamiltonian consists of a drive on qubit A, a non-reciprocal interaction between A and B, and finite excited state lifetimes for both qubits.
In this regime, the effective population in $|11\rangle$ is of the order $(\Omega/\gamma)^2$ and can be neglected, leading to a pure fixed point (steady state) of the effective Hamiltonian. The state reads
\begin{equation}
    |\psi(t \rightarrow \infty)\rangle = |00\rangle - \frac{i2\Omega}{\gamma}|10\rangle + \frac{i4\Omega}{\gamma}|01\rangle.
\end{equation}

\begin{figure}[t!]
	\centering
	\includegraphics{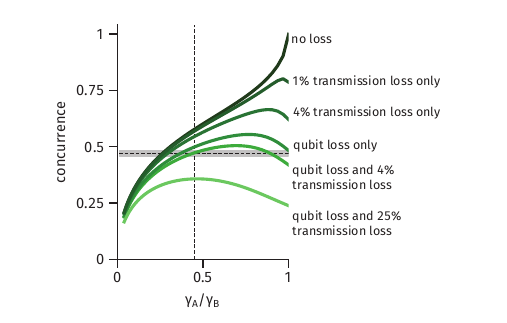}
	\caption{
        \textbf{Effect of loss on maximal concurrence.}
        Simulated maximum concurrence achievable as a function of the ratio of qubit-waveguide coupling strengths. 
        This figure supplements Fig.~\ref{fig:landscape}c by illustrating the impacts of different sources of loss.
        Each line shows the maximum concurrence that can be stabilized with increasing degrees of loss. Darker to lighter: no loss, $1\%$ transmission loss only, $4\%$ transmission loss only, qubit loss only, qubit and $4\%$ transmission loss, qubit loss and $25\%$ transmission loss.
        Black dashed lines: the experimental system investigated in this work; gray shading indicates uncertainty in the concurrence.
        See Table~\ref{tab:transmon parameters} for qubit loss parameters.
        }
	\label{suppfig:effectofloss}
\end{figure}

This state has non-zero overlap with both the singlet and triplet states, and is entangled for any non-zero driving strength.
This perturbative argument does not hold for the drive and coupling strengths used in the experiment. 
However, it provides an intuitive picture for how a finite degree of entanglement may be stabilized by driving only the upstream qubit.

\section{Performance limitations}
\label{app: Expectations and best performance}

The degree of stabilized entanglement is limited by local loss, and the mismatch in qubit-waveguide coupling strengths.
Fig.~\ref{suppfig:effectofloss} illustrates how different degrees of loss reduce the maximum achievable concurrence, as a function of the ratio of qubit-waveguide coupling strengths.

Measurements of the transmon qubits without coupling them to the chiral waveguide yielded intrinsic relaxation times of approximately \qty{40}{\micro\second}, limited by fabrication and enclosure material.
Together, with a transmission efficiency of $96\%$, we predict a maximum achievable concurrence of approximately $\conc \approx 0.5$.
This is obtained after optimizing steady-state concurrence over drive strengths and drive detuning.
Using the CQA scheme, however, the highest attainable concurrence is simulated to be $\conc \approx 0.29$, which is significantly lower than the bound set by loss and coupling mismatch.

Using drive parameters that enforce the `synthetic squeezing' symmetry, the steady-state concurrence ($\conc = 0.471$) is sufficiently close to the bound set by loss and coupling mismatch ($\conc = 0.48$).
This result suggests that higher-fidelity entangled states may be stabilized by improving qubit coherence and reducing transmission loss.
Qubit coherence can be improved by employing techniques to reduce loss due to materials and the enclosure for the chip \cite{ganjam_surpassing_2024}.
Recent work on on-chip chiral interfaces suggests a possible route to reducing transmission loss by replacing microwave circulators~\cite{joshiResonanceFluorescenceChiral2023a, caoParametricallyControlledChiral2024, kannan2023demand}.


%

\end{document}